\documentclass[prd,12pt]{article}
\pdfoutput=1 

\usepackage{diagbox}

\usepackage{tikz}
\usepackage{amsmath,amssymb,graphicx,multirow,xspace,slashed,array,booktabs}
\usepackage[colorlinks=true,urlcolor=blue,anchorcolor=blue,citecolor=blue,filecolor=blue,linkcolor=blue,menucolor=blue,pagecolor=blue]{hyperref}
\usepackage[compress,numbers]{natbib}
\usepackage{placeins}
\usepackage{caption}
\usepackage{subcaption}
\usepackage{booktabs}
\usepackage{blindtext, rotating}
\usepackage{afterpage}
\usepackage{enumitem}
\usepackage{ marvosym }
\usepackage{authblk} 
\usepackage{verbatim}
\usepackage{soul} 
\usepackage[normalem]{ulem}
\usepackage{pifont}
\usepackage{booktabs}
\usepackage{bm}
\usepackage{cleveref}
\usepackage{siunitx}

\usepackage{floatrow}
\newfloatcommand{capbtabbox}{table}[][\FBwidth]

\usepackage[font=footnotesize,labelfont=bf]{caption}

\usepackage{lineno}

\allowdisplaybreaks

\long\def\symbolfootnote[#1]#2{\begingroup%
\def\thefootnote{\fnsymbol{footnote}}\footnote[#1]{#2}\endgroup}

\allowdisplaybreaks

\makeatletter
	
	\@addtoreset{equation}{section}
\makeatother

\newcommand{\newc}{\newcommand}
\newc{\gsim}{\lower.7ex\hbox{$\;\stackrel{\textstyle>}{\sim}\;$}}
\newc{\lsim}{\lower.7ex\hbox{$\;\stackrel{\textstyle<}{\sim}\;$}}
\newc{\gev}{\,{\rm GeV}}
\newc{\mev}{\,{\rm MeV}}
\newc{\ev}{\,{\rm eV}}
\newc{\kev}{\,{\rm keV}}
\newc{\tev}{\,{\rm TeV}}
\newc{\MHT}{$H_T^{\text{miss}}$}
\newc{\MET}{$\slashed{E}_T$}
\newc{\MTT}{$M_{T2}$}

\def\ln{\mathop{\rm ln}}

\newc{\mz}{M_Z}
\newc{\mpl}{M_*}
\newc{\mw}{m_{\rm weak}}
\newc{\nr}[1]{N^c_R{}_{#1}}

\usepackage{accents}
\newlength{\dhatheight}

\def\beq{\begin{equation}}
\def\eeq{\end{equation}}
\newcommand{\bea}{\begin{eqnarray}\begin{aligned}}
\newcommand{\eea}{\end{aligned}\end{eqnarray}}
\def\bitem{\begin{itemize}}
\def\eitem{\end{itemize}}

\begin{document}
\baselineskip 0.6cm

\begin{titlepage}

\vspace*{-0.5cm}

\thispagestyle{empty}

\begin{center}

\vskip 0.7cm

{\Huge \bf
Aiming for Tops of ALPs \\[1ex] with a Muon Collider 
}

\vskip 0.7cm

\vskip 0.7cm
{\large So Chigusa$^{1,2}$, Sudhakantha Girmohanta$^{3,4}$, \\[1ex]
Yuichiro Nakai$^{3,4}$ and Yufei Zhang$^{3,4}$}
\vskip 1.0cm
{\it
$^1$Berkeley Center for Theoretical Physics, Department of Physics, \\
University of California, Berkeley, CA 94720, USA\\
$^2$Theoretical Physics Group, \\
Lawrence Berkeley National Laboratory, Berkeley, CA 94720, USA \\
$^3$Tsung-Dao Lee Institute, Shanghai Jiao Tong University, \\
520 Shengrong Road, Shanghai 201210, China \\
$^4$School of Physics and Astronomy, Shanghai Jiao Tong University, \\
800 Dongchuan Road, Shanghai 200240, China}
\vskip 1.0cm

\end{center}

\vskip 0.5cm

\begin{abstract}

Future muon colliders with center-of-mass energy of $\mathcal{O}(1-10)$ TeV can provide a clean high-energy environment
with advantages in searches for TeV-scale axion-like particles (ALPs),
pseudo-Nambu–Goldstone bosons associated with spontaneously broken global symmetries, which are widely predicted
in physics beyond the Standard Model (SM).
We exploit ALP couplings to SM fermions, and 
guided by unitarity constraints, build a search strategy focusing on the ALP decay to top quark pairs at muon colliders.
It is found that a large parameter space of TeV-scale ALPs with TeV-scale decay constants
can be probed by utilizing the ALP-top quark coupling.

\end{abstract}

\flushbottom

\end{titlepage}


\section{Introduction}\label{intro}

Axion-like particles (ALPs) are pseudo-Nambu–Goldstone bosons (pNGBs)
associated with spontaneously broken global symmetries.
The exploration of such particles is inspired by the QCD axion
\cite{Peccei:1977hh,Weinberg:1977ma,Wilczek:1977pj}
to solve the strong CP problem
\cite{Baker:2006ts,Pendlebury:2015lrz},
but unlike the QCD axion, their masses and decay constants are independent parameters and can vary over a wide range.
ALPs are predicted by many extensions of the Standard Model (SM)
and naturally arise in string theory
\cite{Svrcek:2006yi,Arvanitaki:2009fg}.
They have various phenomenological, cosmological, and astrophysical implications
that have been extensively studied
(for reviews, see $e.g.$ refs.~\cite{Beacham:2019nyx,Choi:2020rgn}).
Therefore, ALPs are considered to be one of the most important targets in modern particle physics,
and their discoveries will make tremendous progress in the understanding of our universe or multiverse.

Among a wide range of ALP models, our focus in the present paper is on ALPs whose masses
and decay constants are around the TeV scale.
Such TeV-scale ALPs are well-motivated by solutions to the so-called axion quality problem
that is a challenge for the Peccei-Quinn (PQ) mechanism to solve the strong CP problem
\cite{Holman:1992us,Kamionkowski:1992mf,Barr:1992qq,Ghigna:1992iv,Carpenter:2009zs}.
A spontaneously broken global symmetry, resulting in the QCD axion, must be very exact,
otherwise, the axion potential would be minimized at an inappropriate value to cancel the strong CP phase.
However, any global symmetry is expected to be explicitly broken by quantum gravity effects.
Among various possibilities, one approach to the axion quality problem is to increase the axion mass
without destroying the PQ solution to the strong CP problem
\cite{Dimopoulos:1979pp,Holdom:1982ex,Dine:1986bg,Flynn:1987rs,Rubakov:1997vp,Berezhiani:2000gh,Fukuda:2015ana,Gherghetta:2016fhp,Dimopoulos:2016lvn,Agrawal:2017ksf,Gaillard:2018xgk,Hook:2019qoh,Gherghetta:2020keg,Gherghetta:2020ofz}.
If the axion is heavy enough, quantum gravity effects are no longer problematic,
and the axion decay constant can be set to the TeV scale.
Another interesting possibility is to consider a warped extra dimension model with three 3-branes
where the PQ symmetry is spontaneously broken on the intermediate 3-brane
and the electroweak symmetry is broken by Higgs fields localized on the IR brane
\cite{Lee:2021slp}
(for recent developments of multi-brane models, see $e.g.$ refs.~\cite{Lee:2021wau,Girmohanta:2023sjv}).
This model can address the electroweak naturalness problem
and the strong CP problem with a high-quality axion at the same time.
A unique prediction of the model is the presence of Kaluza-Klein resonances
associated with the invisible QCD axion, which are TeV-scale ALPs with TeV-scale decay constants.
Since the parameter space of the ALP mass and decay constant that the Large Hadron Collider (LHC) can explore
is limited, new high-energy machines are needed to probe such TeV-scale ALPs.

A future muon collider with center-of-mass (CM) energy of $\mathcal{O}(1-10)$ TeV can furnish a clean high-energy environment
with advantages in searches for TeV-scale ALPs.
Since the muon is much heavier than the electron and loses much less energy by the synchrotron radiation,
a multi-TeV muon beam with a high luminosity can be realized in a relatively small circular ring. 
A muon collider is also able to provide clean and precise collisions of point-like particles,
unlike hadron colliders that involve composite particles.
Due to the short muon lifetime of $\SI{2.2}{\micro s}$ at rest, it is required to produce a large number of muons, cool them down to reduce their emittance, and accelerate them quickly before they decay.
Recent technological developments
\cite{MICE:2019jkl,Delahaye:2019omf,Bartosik:2020xwr}
make it possible to solve those issues and move forward to the realization of a high-energy muon collider.
Therefore, a muon collider has currently received strong support and interest from the particle physics community. For recent studies on the physics reach for a future muon collider,
see refs.~\cite{AlAli:2021let,Black:2022cth,Aime:2022flm,Hamada:2022mua} and references therein.

Depending on the structure of ALP couplings to SM particles,
ALP production and decay at a high-energy muon collider can vary significantly.
The authors of refs.~\cite{Han:2022mzp, Bao:2022onq} have initiated studies on searches for TeV-scale ALPs
coupling to the electroweak gauge bosons,
demonstrating that a muon collider can substantially expand the ALP mass coverage.
In the present paper, we focus on ALP couplings to SM fermions,
which are naturally expected to be present in the axion effective theory.
Since the ALP-fermion coupling is proportional to the fermion mass,
TeV-scale ALPs predominantly decay into top quark pairs in this scenario.
We establish a search strategy for TeV-scale ALPs accordingly;
it is demonstrated that the number of jets and reconstructed top candidates can serve as filters
to distinguish the signal events we are interested in.
We elaborate on the detailed procedure for reducing the principal backgrounds, including beam-induced fake jets.
Furthermore, we reconstruct the dijet invariant mass distribution for the outgoing top quark pairs.
Searching for a peak structure in the distribution helps to improve the sensitivity,
and also provides information about the ALP mass.
We employ a likelihood analysis to evaluate the significance of the peak observed
in the dijet invariant mass distribution for numerous ALP masses.
It is shown that a large parameter space can be probed by utilizing the ALP-top quark coupling.

The rest of the paper is organized as follows.
Section~\ref{sec:ALPs} introduces the ALP model, presents our setup of ALP couplings to SM particles, and
investigates the ALP signal at a muon collider.
Then, in section~\ref{analysis}, we explain our collider simulations,
discuss the event selection strategy, and explain our statistical treatment.
The results are presented in section~\ref{results}.
Section~\ref{conclusion} is devoted to conclusions and discussions.

\section{Setup}
\label{sec:ALPs}

This section establishes the effective field theoretic framework for the ALP Lagrangian
while elucidating the approximate unitarity constraints associated with its couplings.
Subsequently, we consider the production channels within a prospective muon collider,
focusing on the direct couplings of ALPs with fermions.

\subsection{The effective field theory framework}\label{setup}

We define an effective field theory (EFT) by extending the SM with a CP-odd pNGB $a$, which acts as the ALP and are associated with the breaking of a $U(1)$ global symmetry at a scale $f_a$.
We will work in the SM effective field theory framework,
where the electroweak symmetry is realized linearly.
The $U(1)$ breaking is assumed to be above the scale of the electroweak symmetry breaking,
i.e., $f_a > 250$ GeV so that the effective operators are full $SU(3)_c \otimes SU(2)_L \otimes U(1)_Y$ invariant and
ALP couplings to the EW gauge bosons become important for a future muon collider
where the vector boson fusion (VBF) processes can be substantial.
The Lagrangian can be expanded in powers of $a/f_a$, to the leading order,
		\begin{equation}
			{\cal L}_{\rm eff} = {\cal L}_{\rm SM} + \frac{1}{2} (\partial_\mu a) (\partial^\mu a) - \frac{1}{2} m_a^2 a^2 + \delta {\cal L}^{(a)} \ .
			\label{Leff_eqn}
		\end{equation}
Here, ${\cal L}_{\rm SM}$ denotes the ordinary SM Lagrangian, $m_a$ is the ALP mass, and
	\begin{equation}
 \label{deltaL}
		\delta {\cal L}^{(a)} = c_{\widetilde{W}} {\cal A}_{\widetilde{W}} + c_{\widetilde{B}} {\cal A}_{\widetilde{B}} + c_{\widetilde{G}} {\cal A}_{\widetilde{G}} + c_{a \Phi}  O_{a\Phi}^{\psi} \ ,
	\end{equation}
with
	\begin{align}
		O_{a\Phi}^{\psi} &=i \frac{a}{f_a} \sum_{\psi=Q, L} \bar\psi_L \mathbf{Y_\psi} \mathbf{\Phi} \sigma_3 \psi_R
  + {\rm h.c.} \ ,
		\label{eq:OaPhi}
	\end{align}
where $Q_{R} \equiv (u_R, d_R)$, $L_R \equiv (0, e_R)$, $\sigma_3$ acts on the weak isospin,
${\mathbf{Y}_Q} \equiv {\rm diag} (Y_U, Y_D)$, $\mathbf{Y}_L \equiv {\rm diag} (0, Y_E)$,
and $\mathbf{\Phi} \equiv (\tilde{\Phi}, \Phi)$ with $\Phi$ being the SM Higgs doublet
and $\tilde{\Phi} = i \sigma_2 \Phi^\ast$. 
Furthermore, $Y_U$, $Y_D$, and $Y_E$ represent the Yukawa coupling matrices in the flavor space
for the up-type quarks, down-type quarks, and charged leptons, respectively.
In eq.~\eqref{deltaL}, $c_{\widetilde{W}, \widetilde{B}, \widetilde{G}, a \Phi}$ are dimensionless coefficients and we have used the notation,
	\begin{equation}
		{\cal A}_{\widetilde{X}} = -X_{\mu \nu} \widetilde{X}^{\mu \nu} \frac{a}{f_a} \ ,
	\end{equation}
for a gauge field $X \in \{B, W, G\}$ with the field strength tensor $X^{\mu \nu}$ and the dual tensor $\widetilde{X}^{\mu \nu} = \epsilon^{\mu \nu \rho \sigma} X_{\rho \sigma}$.
For the corresponding Feynman rules, see Appendix B of ref.~\cite{Brivio:2017ije}.

ALP couplings to the SM vector bosons have played a central role in building ALP search strategies
at a future muon collider~\cite{ Han:2022mzp,Bao:2022onq}.
On the other hand, the couplings of the ALP to SM fermions were not considered.
However, these couplings are natural, and not prohibited by any symmetry principles.
Furthermore, the couplings are constrained less stringently by the unitarity bound from partial wave analysis~\cite{Brivio:2021fog}.
The helicity amplitude for fermion scattering processes such as $u_k u_{k'} \to Z a$,  where $k, k'$ are flavor indices,
places partial-wave unitarity bounds on the direct fermion coupling, the most restrictive of which is~\cite{Brivio:2021fog} 
\begin{equation}
    \left|c_{a\Phi}\right| \lesssim 6 \left(\frac{f_a}{\rm TeV}\right) \left( \frac{5 \ \rm TeV}{\sqrt{s}}\right) \ .
    \label{CaPhi_bound}
\end{equation}
The couplings to vector bosons ($c_{\tilde{G}}, c_{\tilde{B}}, c_{\tilde{W}}$) are constrained more stringently
from the partial wave unitarity bounds of the scatterings of vector bosons via the ALP (e.g., $ZZ \to ZZ$).

\begin{figure}[!t]
	\includegraphics[width=0.7\textwidth]{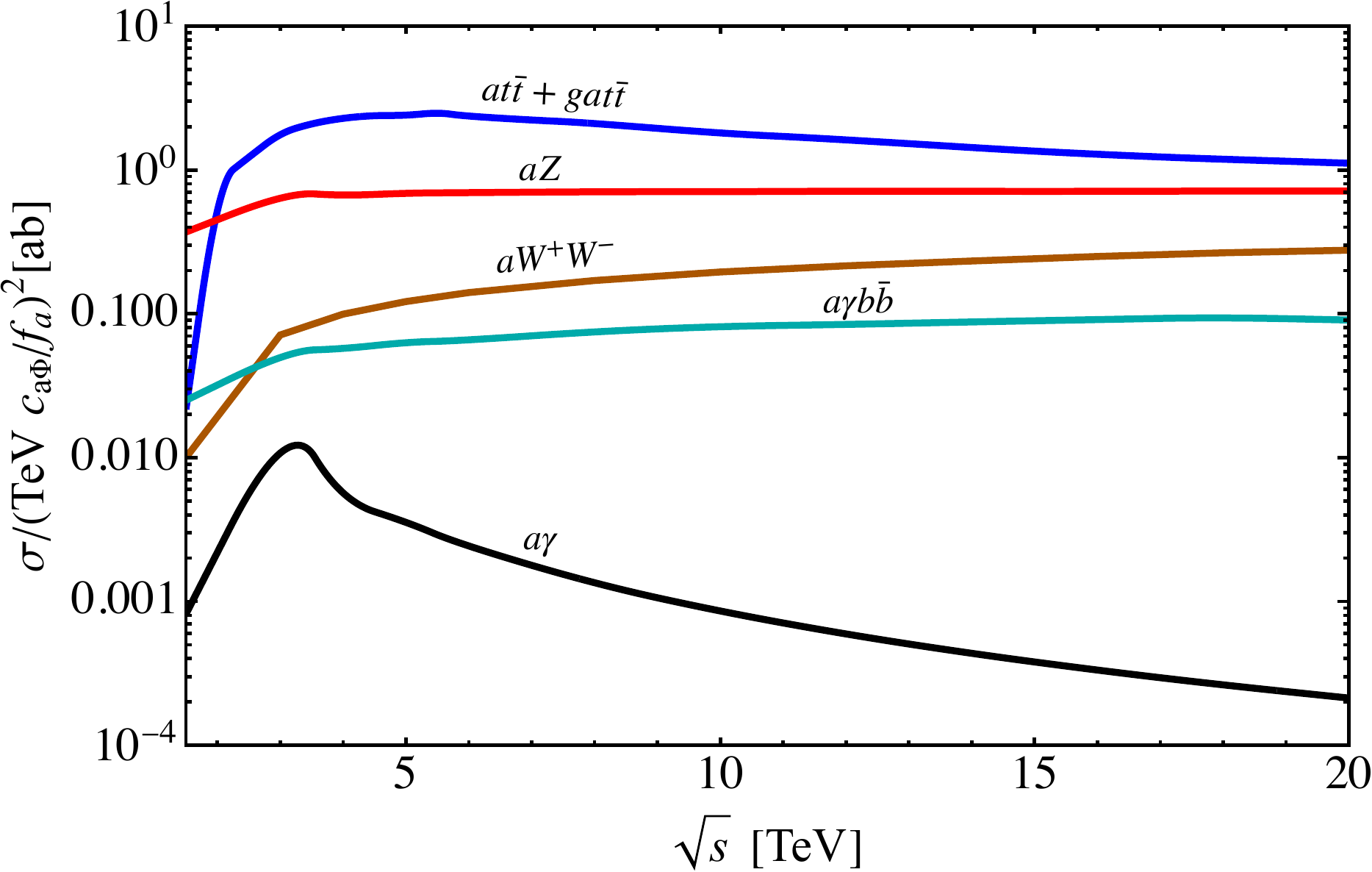}
	\caption{
    Variation of ALP production cross-sections with CM energy for different channels.
    We only have non-zero $c_{a\Phi}$, where all the other couplings are set to zero.
    The plot demonstrates the best channel for the search, namely $\mu^+ \mu^- \to a t \bar{t}$. 
	}
 \label{fig:sigeff}
\end{figure}

In the present paper, we initiate the study of ALP search strategies based on the ALP couplings to the SM fermions
at a muon collider by focusing on a simplified limit in which the ALP-gauge boson couplings are dialed down to zero,
while the effect of the direct couplings to the SM fermions remains.
A generic feature of the ALP couplings to the SM fermions, owing to the shift symmetry,
is that these couplings are proportional to the corresponding fermion masses.
Therefore, an ALP in the TeV mass range dominantly decays to a $t \bar t$ pair.
This fact drives the search strategy we build in the current study.

\subsection{Collider phenomenology}

	\begin{figure}[!t]
		\begin{center}
			\begin{subfigure}{0.32\textwidth}
				\centering
				\includegraphics[width=\textwidth]{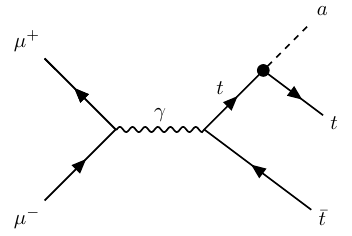}
			\end{subfigure}
			\hfill
			\begin{subfigure}{0.32\textwidth}
				\centering
				\includegraphics[width=\textwidth]{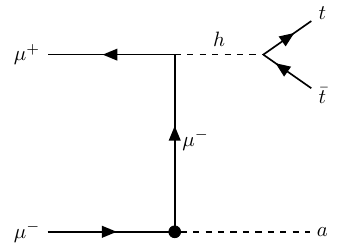}
			\end{subfigure}
        \hfill
			\begin{subfigure}{0.32\textwidth}
				\centering
				\includegraphics[width=\textwidth]{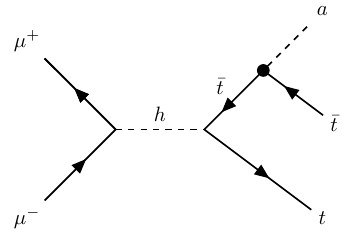}
			\end{subfigure}        
		\end{center}
		\caption{A few representative Feynman diagrams for the signal event $\mu^+ \mu^- \to a t \bar{t}$. The final state ALP predominantly decays to a $t \bar{t}$ pair. } 
		\label{fig:feynman}
	\end{figure}

In figure~\ref{fig:sigeff}, we show the production cross-sections of the ALP for a few representative channels
and their variation with CM energy $\sqrt{s}$ of a muon collider.
As can be seen in the figure, the top-associated production $\mu^-  \mu^+ \to t \bar{t} a (+g)$,
with the matching of jets~\cite{Frixione:2002ik} to avoid double counting,
is the dominant production channel for the ALP
(see figure~\ref{fig:feynman} for a few representative Feynman diagrams).
Note that there are also VBF contributions to the final states including $t\bar{t}a(+g)$,
which has a cross-section as large as $\mathcal{O}(10)\,\mathrm{\%}$ of that of $\mu^-  \mu^+ \to t \bar{t} a (+g)$
in spite of the more severe phase space suppression.
This can be interpreted as a consequence of the large log factor $\ln (s/\Lambda_W^2)$ with $\Lambda_W =  \mathcal{O}(100)\,\mathrm{GeV}$ being the weak scale, which exhibits another role of the muon collider as a vector boson collider.
In addition, we have checked that the processes with more than one gluon emission only give $\lesssim 10\,\mathrm{\%}$ correction to the cross-section for $\sqrt{s}=5\,\mathrm{TeV}$.
Note, however, that these additional QCD processes can be important for larger $\sqrt{s}$.
In the present work, we focus only on the leading production channel $at\bar{t}(+g)$.

Since the ALP couples to the SM fermion with a coupling strength proportional to the Yukawa coupling as shown in \cref{eq:OaPhi},
the ALP most likely decays into a $t \bar{t}$ pair dominantly, when it is kinematically allowed.
Moreover, the ALP decays promptly for models with large fermionic couplings of our interest.
For example, an illustrative choice for the parameter set, $m_a=1\,\mathrm{TeV}$ and $|c_{a\Phi}|/f_a=1/(50\,\mathrm{TeV})$, leads to the branching ratio $\mathrm{Br}\,(a\to t\bar{t}) > 99.9\,\%$ with the partial decay width $\Gamma(a\to t\bar{t}) \sim 10^{-3}\,\mathrm{GeV}$, which corresponds to the typical flight length $c\tau_a \sim 200\,\mathrm{fm}$ with $c$ and $\tau_a$ being the speed of light and the ALP lifetime, respectively.
Thus, combined with the production process $at\bar{t}$ of our interest, we need to consider a way to look for events with four top quarks.

As explained in detail below, we focus on the hadronic decays of top quarks and identify the boosted top-jet candidate by using the reconstructed jet mass. 
This method effectively selects the top-initiated jets, in which all three hadronic decay products of a single top are within the jet cone.
The method works well because of the high collision energy of a muon collider, for example, with $\sqrt{s}=\SI{5}{TeV}$; in contrast, for four-top events at the LHC with $\sqrt{s}=\SI{14}{TeV}$, the boosted top-jet reconstruction is not very efficient since the parton-level collision energy is typically $\sqrt{s'}\lesssim$ a few $\si{TeV}$, and the analysis with leptonic decays is more preferable~\cite{ATLAS:2021kqb, ATLAS:2023ajo, CMS:2023ftu}. In our analysis, we require at least three top-jet candidates in the event. Therefore, most of the background events come from the $t\bar{t}t\bar{t}$ production within the SM, which does not have an axion as an intermediate state. As we will show later in \cref{sec:event_selection}, some background events also come from $t\bar{t}+$jets channels, where a non-top-jet can be misidentified as a top candidate. For reference, \cref{tab:cross_sections} shows the cross-sections of processes relevant to our analysis for the benchmark model of \cref{benchmark_table}.

\begin{table}
\begin{center}
\begin{tabular}{|c|c|c|c|c|} \hline
$c_{\tilde{W}}$ & $c_{\tilde{B}}$ & $c_{\tilde{G}}$ & $|c_{a\Phi}|/f_a$ [TeV$^{-1}$] & $m_a$ [TeV]\\
\hline
$0$ & $0$ & $0$ & $6$ & $1$ \\
\hline
\end{tabular}
\end{center}
\vspace{-0.4cm}
  \caption{A benchmark model parameters we use in our analysis.
	The coupling value is the maximum of the allowed values for $\sqrt{s}=\SI{5}{TeV}$ from the unitarity bound~\eqref{CaPhi_bound}.}
  \label{benchmark_table}
\end{table}

\section{Analysis}
\label{analysis}

We now describe the event generation process for the signal together with some comments on the reduction of fake jets sourced from the beam-induced backgrounds (BIBs).
Then we outline our event selection strategy, considering various potential backgrounds that may arise there.
Principal backgrounds are identified and the way to reduce them is explained.
Finally, we outline the statistical method to estimate the signal significance.

\subsection{Event generation}

In the following discussion, we choose $\sqrt{s}=5\,\mathrm{TeV}$, $\mathcal{L}=\SI{100}{ab^{-1}}$,
and benchmark parameters shown in table~\ref{benchmark_table} for an illustrative purpose.
The events are first generated by MadGraph5~\cite{Alwall:2014hca}, which is further processed by Pythia8~\cite{Bierlich:2022pfr} for matching of jets, showering and hadronization.
The jet clustering is performed by FastJet~\cite{Cacciari:2011ma} with the default clustering parameter $R=0.5$ and the detector simulation by Delphes~\cite{deFavereau:2013fsa} using the muon collider detector card~\cite{delphesTalk}.
To generate the signal events with an axion in an intermediate state, the model Lagrangian is implemented in FeynRules~\cite{Alloul:2013bka}, which then generates a model file compatible with MadGraph5.

\begin{table*}[!t]\centering
	\setlength{\tabcolsep}{3mm}
	\begin{tabular}{|c||c|c|c|c|c|} \hline
		&$ t \bar{t} a$ & $ t \bar{t} t \bar{t}$ & $ t \bar{t} W^+ W^-$ & $ t \bar{t} h$ & $ t \bar{t}  Z$
		\\ \hline
		$\sigma$[ab] & 90 & 23 & 465 &  324 & 158 \\ \hline
		
	\end{tabular}
 \vspace{0.1cm}
	\caption{
	Cross-sections for the signal and principal backgrounds for the chosen benchmark parameters in table~\ref{benchmark_table}. }
 \label{tab:cross_sections}
\end{table*}

As the muon lifetime is $\SI{2.2}{\micro s}$~\cite{pdg:2022} in its rest frame,
the decay products of the muon beam and their secondary interactions give rise to the BIBs
that pose a major challenge in distinguishing true jets from the beam-induced fake jets.
The fake jets have different characteristics from the true jets,
which can be exploited to separate them from each other.
For example, most of the BIBs have low transverse momenta, asynchronous time of arrival, displaced origin,
and high absolute value of pseudo-rapidity.
Therefore, the following filtering procedure during the jet reconstruction is shown
to efficiently reduce the effect of the BIBs~\cite{MuonCollider:2022ded}:

\begin{itemize}
	\item Tracks are filtered by requiring a number of Vertex Detector hits greater than $3$ and a number of Inner Tracker hits greater than $2$.
	\item The normalized calorimeter hit time $t_N$ should be within a time window $|t_N| < \SI{250}{ps}$. Here, $t_N \equiv t - t_0 - c D$, with $t$ being the absolute hit time, $t_0$ the collision time, and $D$ the hit distance from the collision point.
	\item Each jet should contain at least one track.
\end{itemize}
Furthermore, the following kinematic cuts are useful to reduce the number of fake jets:
\begin{itemize}
	\item The pseudo-rapidity cut $|\eta| < 1.5$ to reject most of the fake jets that lie along the beam axis.
	\item The transverse momentum cut with $p_{\rm T} > \SI{50}{GeV}$.
\end{itemize}

According to ref.~\cite{MuonCollider:2022ded}, the filtering procedure has only a mild effect on
the real jets with $|\eta| < 1.5$ and $p_{\rm T} > \SI{50}{GeV}$,
leading to the reconstruction efficiency of roughly $\sim 90\%$,
while combined with the kinematic cuts it reduces the number of fake jets by orders of magnitude.
Thus, in our analysis, we simply assume that the above procedure reduces the effect of the BIBs to a negligible level,
and adopt $90\%$ of real jets that pass the kinematic cuts.\footnote{
Our approach can be regarded as an approximation that the filtering procedure does not drastically change the kinematic distribution of reconstructed jets.
For a more precise treatment, including the effect of the remnant fake jets, we need a full simulation of the BIBs
and detector response, which is beyond the scope of the present paper.
}

\subsection{Event selection}
\label{sec:event_selection}

Our general strategy for the event selection is as follows.
Firstly, we look for four-top events that contain both contributions from the axion signal and the SM processes.
In our analysis, we focus on the hadronic decay of top quarks and identify candidates of boosted top jets using the jet mass information.
Then, we utilize the number of top candidates and the number of jets to trim the events with relatively smaller number of top quarks.
Finally, we search for a resonance peak structure in the dijet invariant mass distribution of top candidates.
Note that the use of boosted top jets under a sufficiently high $\sqrt{s}$ allows us to reconstruct the resonance peak, which is distinct from the search strategy using the leptonic decays in, e.g., ref.~\cite{ATLAS:2021kqb}.

%
\begin{figure}[!t]
	\centering
 	\begin{subfigure}{0.45\textwidth}
		\centering
		\includegraphics[width=\textwidth]{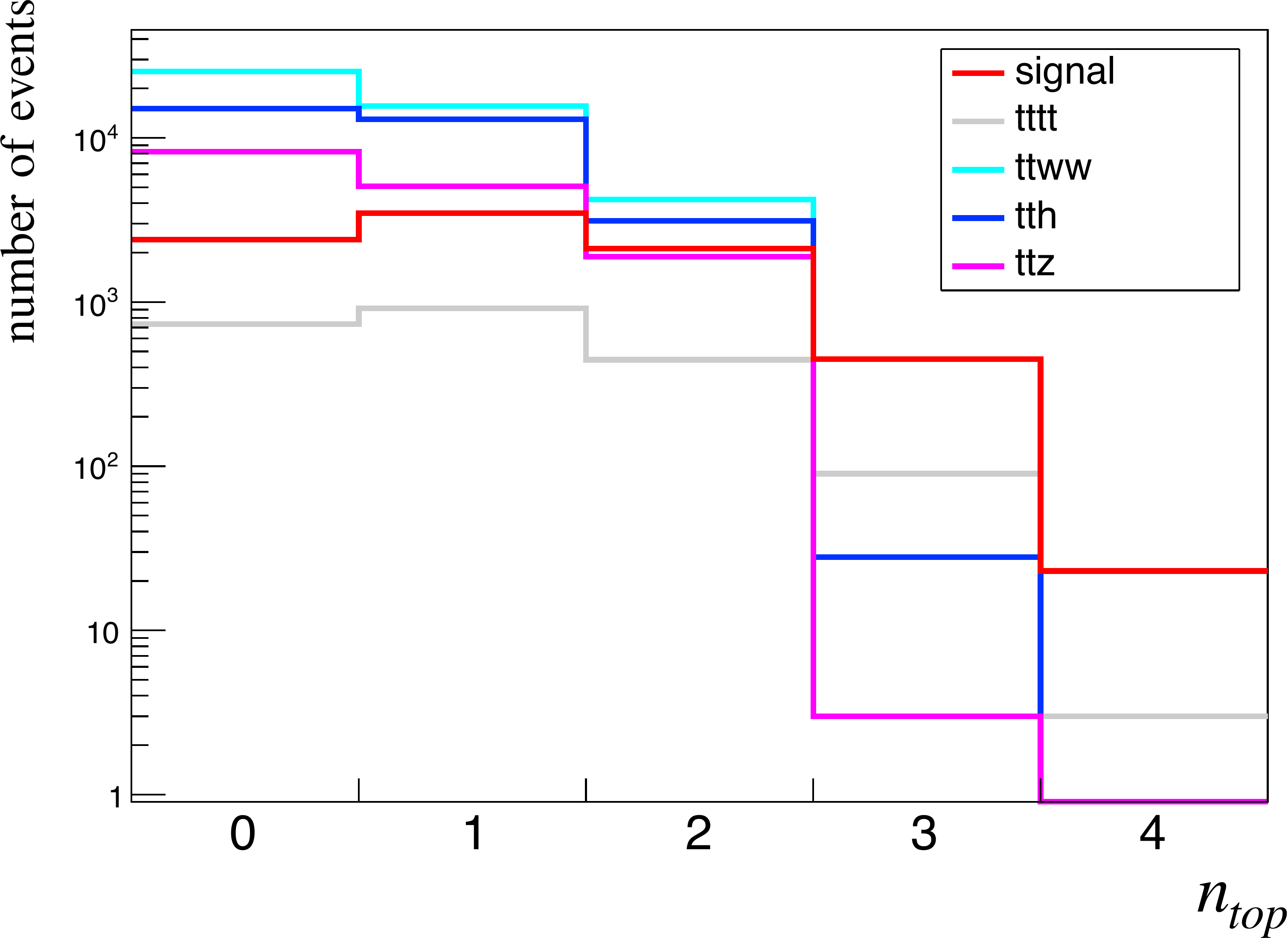}
		\subcaption{$n_{\mathrm{top}}$}
		\label{fig:num_tops}
		\end{subfigure}
  \hspace{0.8cm}
		\begin{subfigure}{0.45\textwidth}
		\centering
		\includegraphics[width=\textwidth]{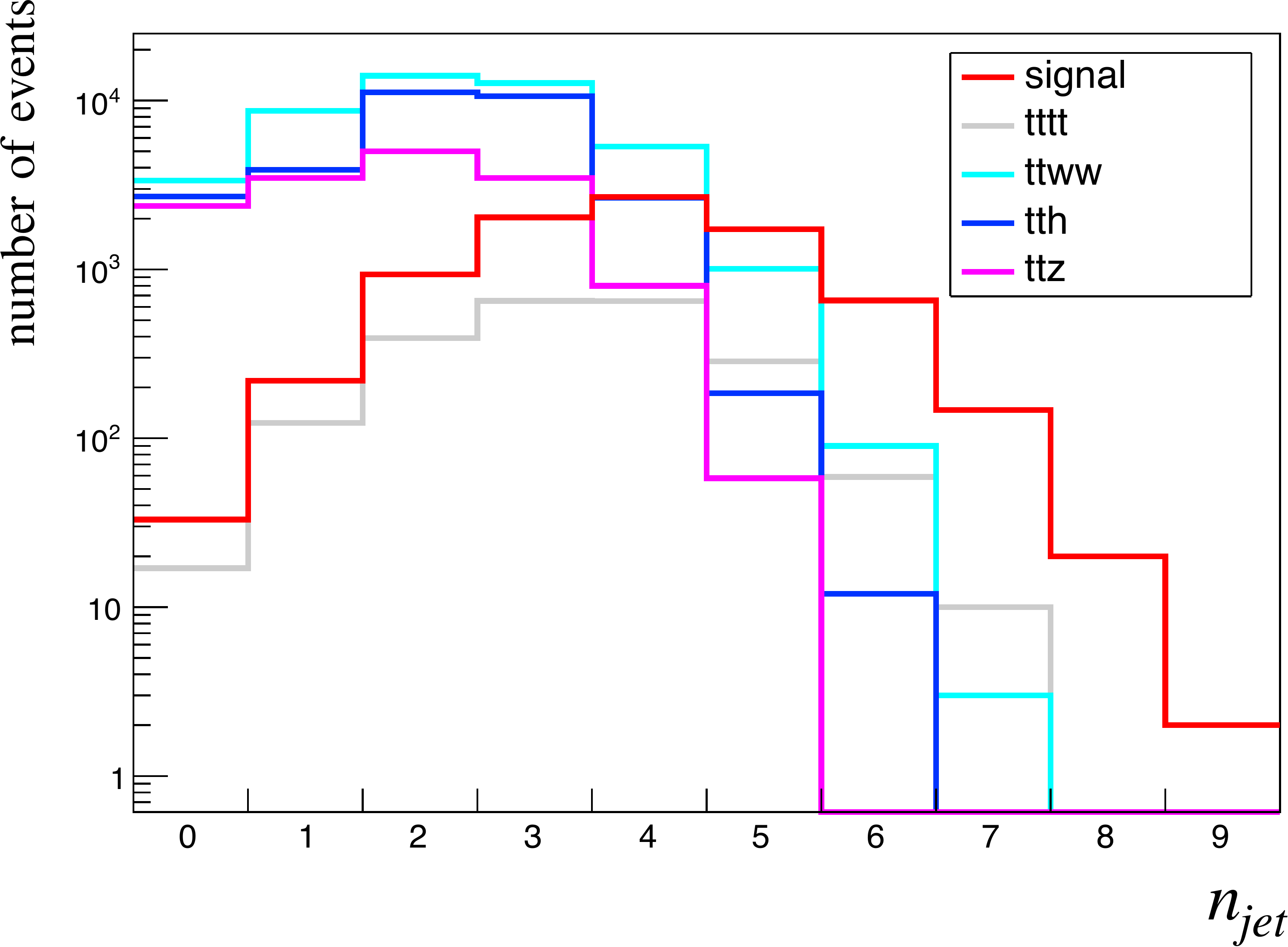}
		\subcaption{$n_{\mathrm{jet}}$}
		\label{fig:num_jets}
		\end{subfigure}
	\caption{
		Distribution of signal and background events with respect to the number of top candidates $n_{\mathrm{top}}$ (left) and jets $n_{\mathrm{jet}}$ (right) defined in the text.
	}
	\label{fig:n_distribution}
\end{figure}
%
To demonstrate what kind of cuts are useful to reduce the number of background events, we first define the number of top candidates $n_{\mathrm{top}}$ and jets $n_{\mathrm{jet}}$ for each event.
Firstly, $n_{\mathrm{jet}}$ corresponds to jets that have passed $|\eta_{\rm jet}|<1.5$ and $p_{\rm T, \rm jet}> \SI{50}{GeV}$ criteria and have been chosen randomly with $90\%$ efficiency as discussed above.
Then, among $n_{\mathrm{jet}}$ jets in an event, we identify those with mass $m_j$ close to the top mass $m_t \simeq \SI{173}{GeV}$~\cite{pdg:2022} as top candidates, which counts to be $n_{\mathrm{top}}$.
We use the mass window $\SI{140}{GeV} < m_j < \SI{220}{GeV}$ for our analysis.
In fig.~\ref{fig:n_distribution}, we show the histograms of $n_{\mathrm{top}}$ (left) and $n_{\mathrm{jet}}$ (right) from the ALP signal as well as some of top-rich SM processes.
From the left panel, it is clear that all the SM processes with only two top quarks tend to have $n_{\mathrm{top}} \leq 2$ as expected, so we can significantly reduce the number of background events with the cut $n_{\mathrm{top}}\geq 3$.
Note that even four-top processes tend to have $n_{\mathrm{top}}<4$ due to the reconstruction efficiency of boosted top jets smaller than $1$.
From the right panel, it can be seen that the number-of-jet cut, e.g., $n_{\mathrm{jet}} \geq 4$ also helps us to reduce backgrounds while maintaining a large part of the signal events.

\begin{table}[!t]\centering
	\setlength{\tabcolsep}{3mm}
	\begin{tabular}{|c||c|c|c|c|c|} \hline
	
		\diagbox{cut}{processes} & $ t \bar{t} a$ & $ t \bar{t} t \bar{t}$ & $ t \bar{t} W^+ W^-$ & $ t \bar{t} h$ & $ t \bar{t}  Z$ 
		\\ \hline
		origin & 9015 & 2285 & 46510 & 32380 & 15790  \\ \hline
        $n_{\rm top}\geq 3$ & 495 & 82 & 37 & 26 & 3  \\ \hline
        $n_{\rm jet}  \geq 4$ & 417 & 65 & 19 & 0 & 0 \\ \hline
	\end{tabular}
 \vspace{2mm}
  	\caption{
		Cutflow for the signal process with benchmark parameters in table~\ref{benchmark_table} and top-rich SM processes for $\sqrt{s}=\SI{5}{TeV}$.
	}
 \label{cutflow_table_5TeV}
\end{table}

Tables~\ref{cutflow_table_5TeV} and \ref{cutflow_table_10TeV} show our cutflow for a simulation of the ALP signal
and top-rich SM processes.\footnote{
It turns out that contributions from the VBF processes are negligibly small after our cutflow for both the signal and background events.
Thus, we neglect these contributions in the tables and the following analysis.
}
Here, we adopt the ALP parameters $m_a=\SI{1}{TeV}$ and $|c_{a\Phi}|/f_a=6$ TeV$^{-1}$ as shown in table~\ref{benchmark_table}.
From \cref{cutflow_table_5TeV,cutflow_table_10TeV}, we can see that the $n_{\mathrm{top}}\geq 3$ cut significantly reduces the number of events from processes with only two top quarks.
The preceding cut $n_{\mathrm{jet}}\geq 4$ further reduces contaminations from $t\bar{t}W^+W^-$ and $t \bar{t} h$ processes, slightly making the signal significance larger.
After all cuts are imposed, most of the events are originated from the four-top processes either through the ALP generation or purely SM processes.
Note that, from comparison between two tables, it is clear that the top-jet reconstruction efficiency becomes higher when we increase $\sqrt{s}$ from $\SI{5}{TeV}$ to $\SI{10}{TeV}$.

From the cutflow, we can order estimate the reconstruction efficiency of a top quark as a top candidate to be of $\mathcal{O}(10)\,\%$, while the misidentification rate of the normal jets to be $\mathcal{O}(1)\,\%$.
Using these values, we estimate how many events survive the cut $n_{\mathrm{top}} \geq 3$, focusing on the SM processes with less than two top quarks in the final state, and conclude that contaminations from these processes are negligible for our analysis.
Thus, the top-rich processes considered in \cref{cutflow_table_5TeV,cutflow_table_10TeV} are the dominant background processes in our analysis. 
Then, we can see from the tables that the number of ALP signal events can be significantly larger than that of the background events after the cutflow, indicating the efficacy of our approach.

\begin{table}[!t]
    \begin{tabular}{|c||c|c|c|c|c|} \hline
	
		\diagbox{cut}{processes} & $ t \bar{t} a$ & $ t \bar{t} t \bar{t}$ & $ t \bar{t} W^+ W^-$ & $ t \bar{t} h$ & $ t \bar{t}  Z$ 
		\\ \hline
		origin & 6625 & 1509 & 34970 & 12300 & 67730  \\ \hline
        $n_{\rm top}\geq 3$ & 784 & 146 & 49 & 9 & 61  \\ \hline
        $n_{\rm jet}  \geq 4$ & 757 & 136 & 24 & 0 & 7 \\ \hline
	\end{tabular}
 \vspace{2mm}
 	\caption{
		Same as \cref*{cutflow_table_5TeV}, but for $\sqrt{s}=\SI{10}{TeV}$.
	}
  \label{cutflow_table_10TeV}
\end{table}

\begin{figure}[t]
	\centering
	\includegraphics[width=0.8\linewidth]{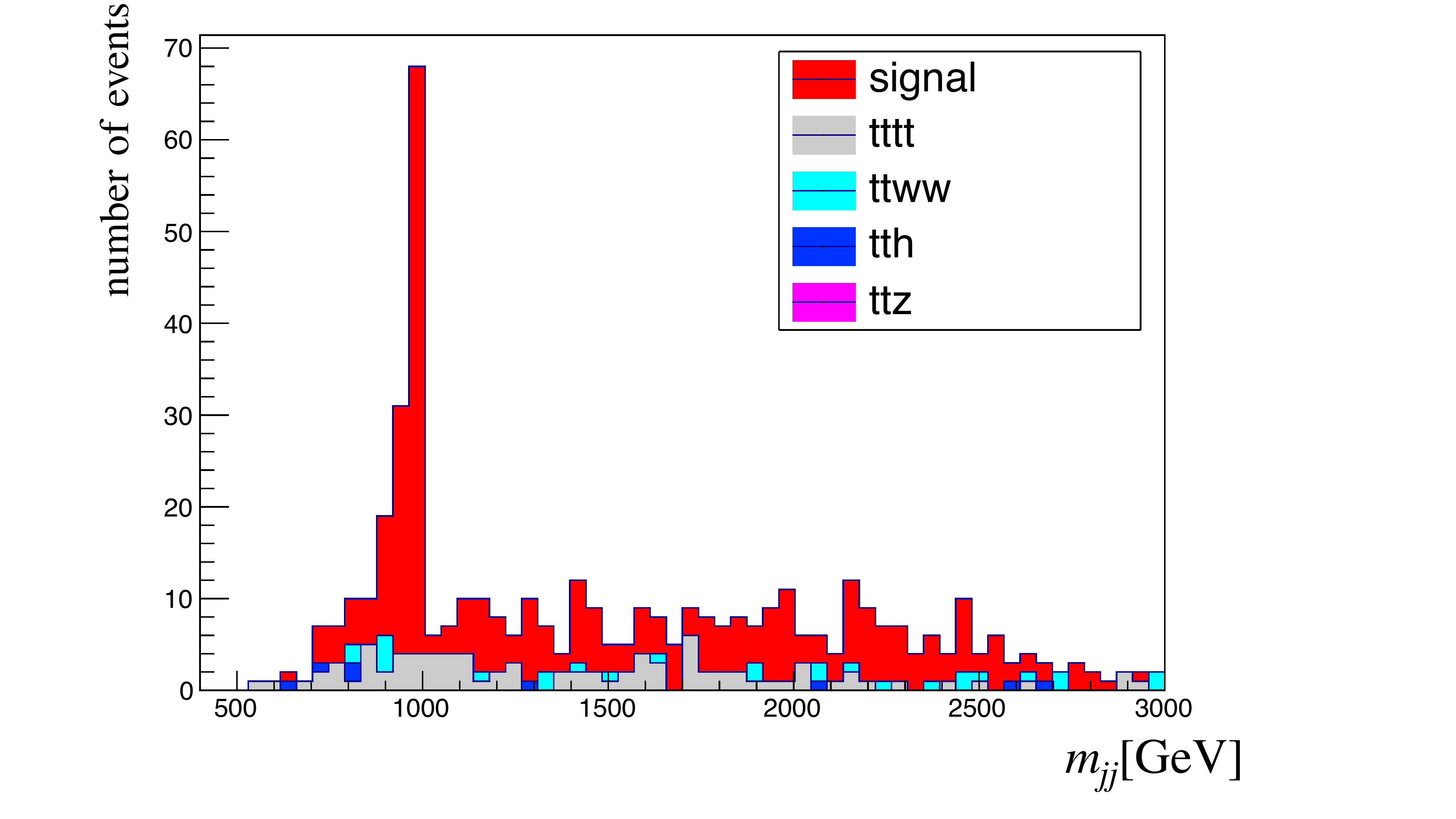}
	\caption{
		Dijet invariant mass distribution of the two top candidates selected as described in the text.
	}
	\label{fig:dijet1TeV}
\end{figure}

Next, we search for a Gaussian peak in the dijet invariant mass distribution of top jets produced by the ALP decay.
Since there are more than two top candidates in the events that passed our cutflow, we need to select two candidates that are most likely sourced from the ALP decay.
We have found that the pair of two top quarks closest to each other in the angular space is a good choice.
More specifically, we define the variable $\Delta R$, which is the angular separation of the two jets calculated as
\begin{equation}
	\Delta R = \sqrt{(\Delta \eta)^2 + (\Delta \phi)^2} \ ,
	\label{deltaR_def:eq}
\end{equation}
where $\Delta \eta$ denotes the pseudo-rapidity difference of the two jets
and $\Delta \phi$ is the azimuthal angle between them.
We calculate $\Delta R$ for every two-jet pair and select the one with the least $\Delta R$.
Fig.~\ref{fig:dijet1TeV} shows the distribution of the dijet invariant mass $m_{jj}$ of the pair of jets selected as such.
We clearly see a peak at $m_{jj}\sim\SI{1}{TeV}$ for the signal events, corresponding to the ALP mass of our choice, while the smooth distribution for $m_{jj} \neq m_a$ is expected to be due to the failure in the choice of two top jets originated from the ALP decay.
We find a systematic shift of the peak position slightly lower than $m_a$, which might come from the missing $p_T$ associated with invisible partons inside jets.
However, we comment that the peak position and height could be used to reconstruct the ALP mass $m_a$ and its coupling with a further understanding of the systematic shift by, e.g., a more profound Monte Carlo simulation of the process.

%


\subsection{Statistical treatment}

In order to quantify the signal significance taking into account the Gaussian peak structure in the dijet mass distribution shown in \cref{fig:dijet1TeV}, we perform fitting of the distribution.
Let $\lambda_i$ be the expected number of events in the $i$-th bin of the histogram.
For the current setup, we decompose $\lambda_i$ as
\begin{align}
	\lambda_i = b_i + s_i
	\;,
	\label{eq:fitfunc_full}
\end{align}
where $b_i$ denotes the smooth distribution that we model with a cubic function,
\begin{align}
	b_i = \sum_{k=0}^{3} \beta_k (m_{jj}^{(i)})^k
	\;,
	\label{eq:fitfunc_bg}
\end{align}
with $m_{jj}^{(i)}$ being a representative value of $m_{jj}$ for the $i$-th bin
and $\beta_k$ ($k=0 \sim 3$) being coefficients, while $s_i$ denotes the peak structure on top of it that we model as
\begin{align}
	s_i = A \exp \left(
		-\frac{\left( m_{jj}^{(i)} - m_0 \right)^2}{2\sigma^2}
	\right)
	\;.
	\label{eq:fitfunc_peak}
\end{align}
Here, $A$, $m_0$, and $\sigma$ are fitting parameters.

Practically, we first prepare the simulation data for each choice of $m_a$ and the coupling.
We fit the data by $\lambda_i$ with fitting parameters being $\beta_k$ ($k=0 \sim 3$), $A$, $m_0$, and $\sigma$.
Let $\tilde{b}_i$ and $\tilde{s}_i$ be the values of $b_i$ and $s_i$, respectively, evaluated with the best-fit values of these parameters.
Defining their combination,
\begin{align}
	\tilde{\lambda}_i(\mu) = \tilde{b}_i + \mu \tilde{s}_i
	\;,
\end{align}
we treat $\lambda_i(\mu=1)$ to be the correct expected number of events for the corresponding ALP model to reduce statistical uncertainties of the simulation.
Note that $\tilde{b}_i$ not only contains a contribution from the SM processes, but also contains a smooth contribution from the process with an intermediate ALP.
This reduces the contribution of the ALP signal to $\tilde{s}_i$, but is useful to make the whole procedure robust to any systematic uncertainties that could modify the di-jet mass distribution smoothly.
Experimentally, $\tilde{b}_i$ and $\tilde{s}_i$ correspond to the fit result of the sideband region with the smooth distribution \eqref{eq:fitfunc_bg} and that of the whole region with the full expression \eqref{eq:fitfunc_full}.

Let $o_i$ be the histogram data obtained from data.
In our analysis, we use $o_i = \tilde{\lambda}_i (\mu=1)$ as a representative data set.
We define the likelihood function as a function of $\mu$ as
\begin{align}
	L(\bm{o}; \mu) = \prod_{i=1}^{N_B} \frac{e^{-\tilde{\lambda}_i(\mu)}\tilde{\lambda}_i(\mu)^{o_i}}{\Gamma(o_i + 1)}
	\;,
\end{align}
with $N_B$ being the number of bins.
We also define a log-likelihood test-statistic,
\begin{align}
	q_0 = -2\ln \frac{L(\bm{o}; \mu=0)}{L(\bm{o}; \mu=1)}
	\label{eq:local_p}
	\;,
\end{align}
where the choice of $\mu=1$ maximizes the denominator.
According to Wilk's theorem~\cite{10.1214/aoms/1177732360}, $q_0$ asymptotically obeys a chi-squared distribution with one degree of freedom.
Thus, we identify $\sqrt{q_0}=5$ as a criteria for the discovery of the ALP model at the $5\sigma$ confidence level.

Note that the procedure described so far gives a local significance of the signal for a fixed axion mass $m_a$ used throughout the analysis.
For the global $p$-value, one can define the test-statistic $q$ by making the $m_a$ dependence of the likelihood function explicit and maximizing the denominator of \cref{eq:local_p} by varying $\mu$ and $m_a$ at the same time.
Since the probability distribution of $q$ is not known a priori, one has to perform a large number of pseudo-experiments to estimate the probability distribution to evaluate the $p$-value corresponding to the expected value of $q$.
This way to estimate a global $p$-value is beyond the scope of the current paper and we leave it as a future task.

\section{Results}
\label{results}

\begin{figure}[ht]
    \centering
    \includegraphics[width=0.7\textwidth]{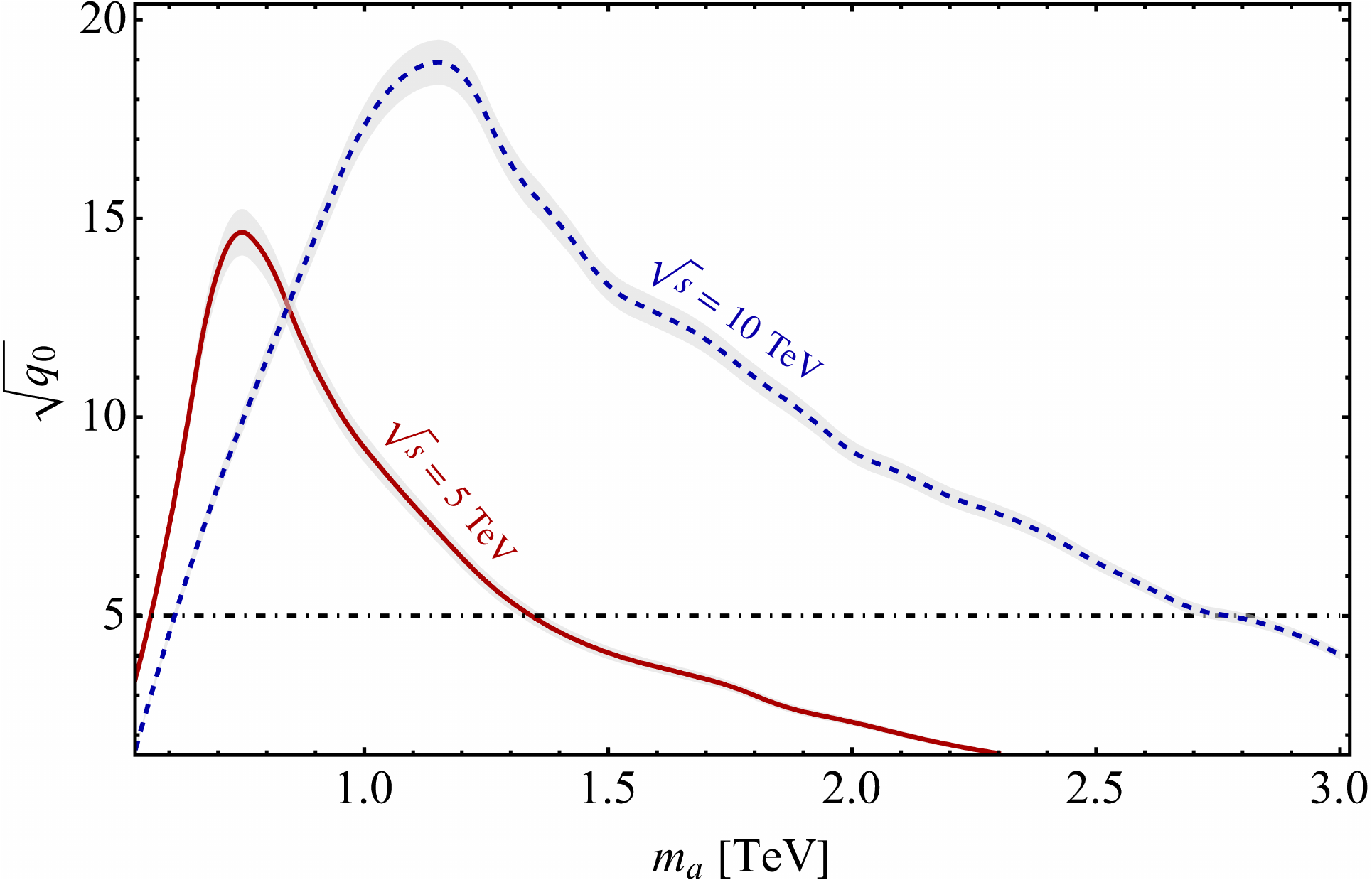}
    \caption{The test-statistic $\sqrt{q_0}$ as a function of the ALP mass $m_a$ for $\sqrt{s}=5$ TeV (red solid), and $\sqrt{s}=10$ TeV (blue dashed) and $\mathcal{L}=\SI{100}{ab^{-1}}$ with other model parameters chosen from table~\ref{benchmark_table}, while the gray band incorporates the uncertainties in the fitting procedure. The black dot-dashed line denotes $\sqrt{q_0}=5$ corresponding to the $5\sigma$ reach.}
    \label{fig:q_vs_ma}
\end{figure}
 
In fig.~\ref{fig:q_vs_ma}, we show the variation of the test-statistic $\sqrt{q_0}$ as a function of different ALP masses while keeping the coupling fixed to $|c_{a\Phi}|/f_a=\SI{6}{TeV^{-1}}$, which is the maximally allowed value by unitarity for $\sqrt{s}=\SI{5}{TeV}$ as shown in table~\ref{benchmark_table}.
The assumed collider setup is $\sqrt{s}=\SI{5}{TeV}$ ($\sqrt{s}=\SI{10}{TeV}$) and $\mathcal{L}=\SI{100}{ab^{-1}}$ for the red solid (blue dashed) line. 
The black dot-dashed line denotes the criteria for the $5\sigma$ discovery, which shows that we can probe up to $m_a \sim 1.3$ TeV ($2.7$ TeV) for $\sqrt{s}=5$ TeV (10 TeV).

The nature of the curve is understood as follows.
Firstly, for $m_a$ barely above the threshold $2 m_t$ of having the $a\to t \bar{t}$ decay channel, top quarks generated from the ALP decay have small momenta in the ALP-rest frame, and they are highly collimated in the lab frame.
Thus, there is a non-negligible probability that hadronic decay products of two top quarks are clustered as a single jet.
We confirm this by verifying the existence of a peak at the ALP mass in the jet mass histogram together with a drastic reduction of $n_{\rm top}=4$ events in our analysis.
This explains the sensitivity decrease for smaller $m_a$, which is more severe for $\sqrt{s}=\SI{10}{TeV}$ with more collimated top jets and more background events.
Note that this sensitivity loss can be well compensated by fitting the jet mass distribution since all the signal events we lose in our analysis should make a more clear peak in the jet mass distribution.

Secondly, for a larger $m_a$ the production cross-section decreases due to the reduction of the available phase space, which basically sets the upper limit on $m_a$ to which our analysis is sensitive. Now, it is useful to understand how this compares with a larger CM energy,
as although we do not gain much in the production cross-section as can be inferred from fig.~\ref{fig:sigeff}, the top-jet reconstruction efficiency increases thanks to higher boost factors.
To verify this, and to compare with the case of $\sqrt{s}=\SI{5}{TeV}$, we set aside the unitarity constraint for the moment, and keep the coupling fixed while increasing CM energy to $\sqrt{s}=\SI{10}{TeV}$. Clearly, due to increased phase space in the larger ALP mass regime, we can probe to higher values of $m_a$, and it is also confirmed that one effectively gets more sensitivity, thanks to the enhanced top-jet reconstruction efficiency.

\begin{figure}[!t]
    \centering
    \includegraphics[width=0.75\textwidth]{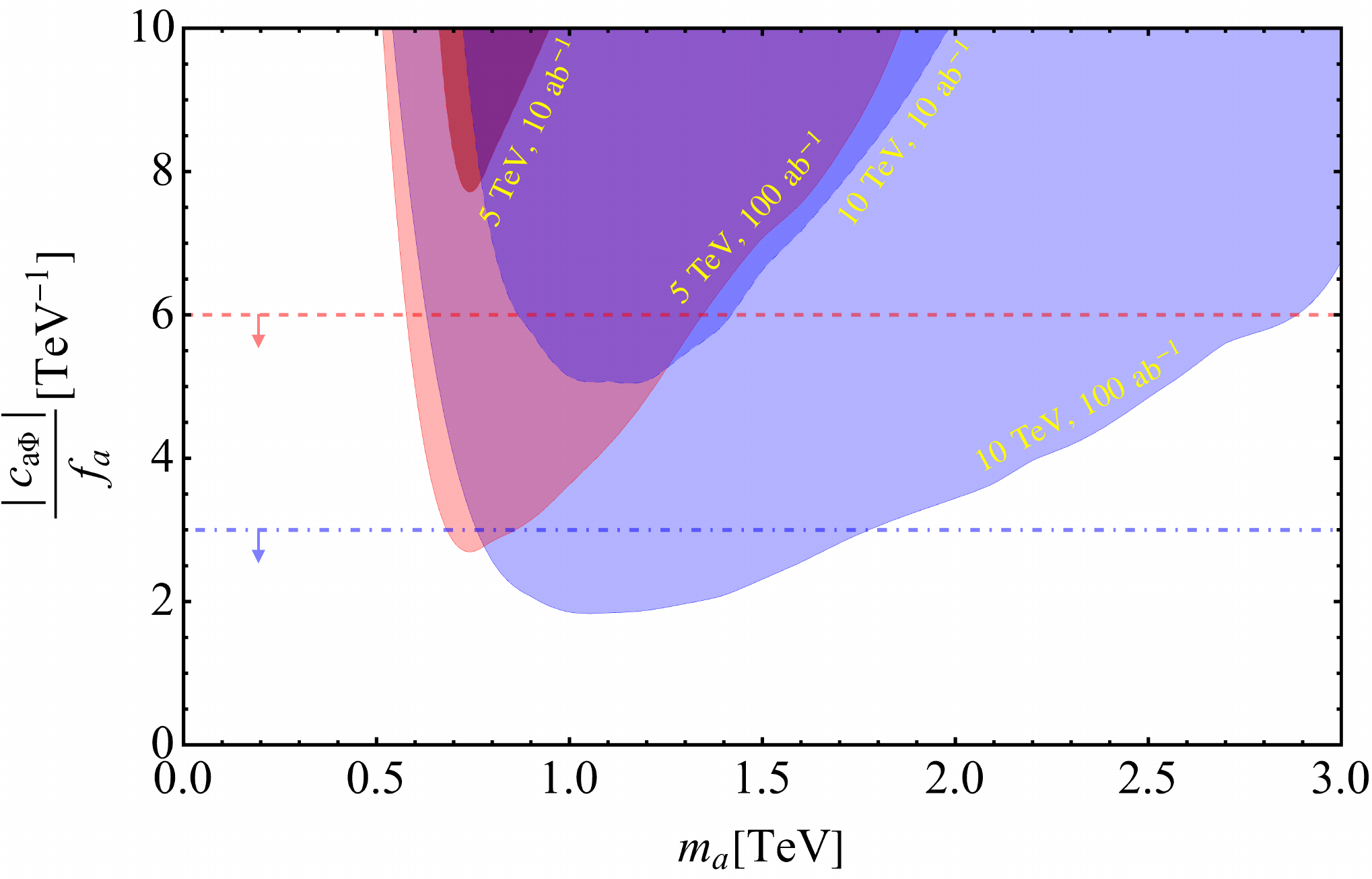}
    \caption{Projected $5\sigma$ reach contours in the normalized coupling $|c_{a\Phi}|/f_a$ and $m_a$ plane using the four-top channel search for two different CM energies, namely $\sqrt{s} = 5$ TeV (red) and $10$ TeV (blue) for integrated luminosities ${\cal L}=10$ {ab}$^{-1}$ and $100$ {ab}$^{-1}$. The approximate unitarity bound~\cite{Brivio:2021fog} is shown as the red dashed (blue dot-dashed) line for $\sqrt{s}=5$ TeV (10 TeV). }
    \label{fig:fvsma}
\end{figure}

Let us now turn to the following question: what is the $5\sigma$ reach for the coupling-mass parameter space, i.e., in the $|c_{a\Phi}|/f_a$-$m_a$ plane? Note that it is difficult to answer this analytically as the test-statistic does not scale with the coupling, and there are non-trivial effects from the background dependence and top-jet reconstruction efficiencies. Therefore, we utilize the following argument, namely that although the test-statistic does not scale with the coupling, the signal cross-section scales as $\left(c_{a\Phi}/f_a\right)^2$. We have verified this numerically within 1\% accuracy. This is because once the ALP is produced on-shell, for $m_a > 2 m_t$, it immediately decays to two top pairs with $\sim 100\%$ branching fraction. Therefore, we can follow the same technique for evaluating $\sqrt{q_0}$, with the same background as before, while re-scaling the signal events, i.e., a reduction in the coupling will reduce the signal peak, and will result in a reduced $\sqrt{q_0}$. Now we view this procedure as a continuous function of the coupling and extract the value of the coupling at which one gets $\sqrt{q_0}=5$ for each fixed value of $m_a$. Hence, this gives the $5\sigma$ reach for the normalized coupling for the chosen ALP mass. We repeat this procedure for all the different events generated for distinct $m_a$, and thereby get the $5\sigma$ projected reach in the coupling-mass plane. This is depicted in fig.~\ref{fig:fvsma} for two different $\sqrt{s}$ and two different $\mathcal{L}$. The estimated unitarity bounds are shown for each of the CM energy as dashed lines, and we can see that one is able to probe a large parameter space including the region where the EFT description is expected to breakdown with possibly the underlying physics signature appearing. For the extra-dimensional KK mode ALP, for example, this could mean that the contribution from the next KK resonance becomes important, while for other UV completion, this could mean the fields that had been integrated out for obtaining the EFT Lagrangian can be excited.
We show both the cases for the integrated luminosities $\SI{10}{ab^{-1}}$ and $\SI{100}{ab^{-1}}$ that help to elucidate the dependence of the parameter space probe as a function of the luminosity.

\section{Discussions}
\label{conclusion}

Future muon colliders with CM energy of $\mathcal{O}(1-10)$ TeV are able to provide a clean high-energy environment
with advantages in searches for TeV-scale ALPs.
We have exploited ALP couplings to the SM fermions, and 
guided by unitarity constraints, built a search strategy focusing on the ALP decay to SM top quark pairs.
We analyzed the extraction of the signal from principal SM backgrounds, and provided a projected sensitivity reach for the parameter space of the ALP.

In our current work, we were agnostic regarding the UV completion,
but our search strategy for the four-top channel that we have elucidated here can be applied to specific models
to find the reach of future collider machines.
Furthermore, the ALP-fermion coupling was considered in isolation to illustrate its effects
on the production and decay channel.
However, in the future, it will be interesting to consider the gauge boson couplings
together with the direct fermion couplings we considered here as this will mimic a more realistic scenario.
Regarding the region $2 m_b < m_a < 2 m_t$, although we do not take up this task in the current work,
note that here the $b \bar{b}$ decay channel for the ALP decay will dominate,
and a similar search strategy could be followed to project a constraint in this regime.
We also note that  High-Luminosity (HL)-LHC will have a comparable or better handle at this mass range,
and one can do a comparative analysis of the efficacy of HL-LHC and a future lepton collider for this regime,
and we relegate it to a future work.\footnote{
Four-top events have been detected recently at both ATLAS and CMS
\cite{ATLAS:2021kqb, ATLAS:2023ajo, CMS:2023ftu}.
}
To conclude, our results show that the ALP-top coupling can be crucial to probe the ALP in four-top channels, as this coupling is natural from a UV theory perspective and can provide a complementary or even better chance at probing it in a future muon collider machine to the more investigated channels
involving gauge boson couplings primarily~\cite{Han:2022mzp, Bao:2022onq}.

\section*{Acknowledgements}

SC is supported by the Director, Office of Science, Office of High Energy Physics of the U.S. Department of Energy under the Contract No. DE-AC02-05CH1123.
YN is supported by Natural Science Foundation of China under grant No. 12150610465.

\appendix

\bibliographystyle{jhep}
\bibliography{rsc_bib}

\end{document}